\documentstyle[12pt]{article}

\def\thefootnote{\fnsymbol{footnote}}
\def\bea{\begin{eqnarray}}
\def\eea{\end{eqnarray}}
\def\be{\begin{equation}}
\def\ee{\end{equation}}

\def\tF{{\tilde F}}

\def\hL{\hat{L}}

\def\[{\left [}
\def\]{\right ]}
\def\({\left (}
\def\){\right )}

\def\pp{\partial}
\def\M{M^*}

\def\S{S^*}

\def\Tr{{\rm Tr}}

\def\G{{\cal G}}

\def\F{{\cal F}}
\def\tcG{{\tilde\G}}
\def\tcF{{\tilde\F}}

\def\L{{\cal L}}

\def\cM{{\cal{M}}}

\def\tG{{\tilde G}}

\def\tM{{\widetilde M}}

\begin{document}

\begin{titlepage}
\begin{center}

\hfill LBNL-41110 \\
\hfill UCB-PTH-97/58 \\
\hfill hep-th/9712103 \\
\hfill \today \\

\vskip .3in
{\large \bf Nonlinear Electromagnetic Self-Duality and Legendre Transformations
 } 
\footnote{Contribution to the Proceedings of the Newton Institute Euroconference
on Duality and Supersymmetric Theories, April, 1997}
\footnote{This work was supported in part by
the Director, Office of Energy Research, Office of High Energy and Nuclear
Physics, Division of High Energy Physics of the U.S. Department of Energy under
Contract DE-AC03-76SF00098 and in part by the National Science Foundation under
grant PHY-95-14797.} \vskip .5in

Mary K. Gaillard {\em and} Bruno Zumino

{\em Physics Department, University of California, and \\
Theoretical Physics Group,  Lawrence Berkeley National Laboratory, 
 Berkeley, California 94720}\\
\vskip .5in

\end{center}

\begin{abstract}

We discuss continuous duality transformations and the properties of classical 
theories with invariant interactions between electromagnetic fields and
matter.  The case of scalar fields is treated in some detail.
Special discrete elements of the continuous group are shown to be related to the
Legendre transformation with respect to the field strengths.

\end{abstract}
\end{titlepage}
\newpage

\renewcommand{\thepage}{\roman{page}}
\setcounter{page}{2}
\mbox{ }

\vskip 1in

\begin{center}
{\bf Disclaimer}
\end{center}

\vskip .2in

\begin{scriptsize}
\begin{quotation}
This document was prepared as an account of work sponsored by the United
States Government. While this document is believed to contain correct 
information, neither the United States Government nor any agency
thereof, nor The Regents of the University of California, nor any of their
employees, makes any warranty, express or implied, or assumes any legal
liability or responsibility for the accuracy, completeness, or usefulness
of any information, apparatus, product, or process disclosed, or represents
that its use would not infringe privately owned rights.  Reference herein
to any specific commercial products process, or service by its trade name,
trademark, manufacturer, or otherwise, does not necessarily constitute or
imply its endorsement, recommendation, or favoring by the United States
Government or any agency thereof, or The Regents of the University of
California.  The views and opinions of authors expressed herein do not
necessarily state or reflect those of the United States Government or any
agency thereof, or The Regents of the University of California.
\end{quotation}
\end{scriptsize}

\vskip 2in

\begin{center}
\begin{small}
{\it Lawrence Berkeley Laboratory is an equal opportunity employer.}
\end{small}
\end{center}

\newpage
\renewcommand{\thepage}{\arabic{page}}
\setcounter{page}{1}
\def\thefootnote{\arabic{footnote}}
\setcounter{footnote}{0}
\renewcommand{\theequation}{\arabic{section}.\arabic{equation}}

\section{Duality rotations in four dimensions}
\setcounter{equation}{0}
The invariance of Maxwell's equations under ``duality rotations'' has been known
for a long time.  In relativistic notation these are rotations of the
electromagnetic field strength $F_{\mu\nu}$ into its dual, which is defined by
\be \tF_{\mu\nu} = {1\over2}\epsilon_{\mu\nu\lambda\sigma}F^{\lambda\sigma},
\quad {\tilde \tF_{\mu\nu}} = -F_{\mu\nu}.\label{g1} \ee
This invariance can be extended to electromagnetic fields in interaction with
the gravitational field, which does not transform under duality.  It  is present
in ungauged extended supergravity theories, in which case it generalizes to a
nonabelian group~\cite{fsz}.  In~\cite{gz,bz} we studied the most general
situation in which classical duality invariance of this type can occur.  More
recently~\cite{gib} the duality invariance of the Born-Infeld theory, suitably
coupled to the dilaton and axion~\cite{gib2}, has been studied in considerable
detail.  In the present note we will show that most of the results 
of~\cite{gib,gib2} follow quite easily from our earlier general discussion.  
We shall also present some new results.

    We begin by recalling and completing some basic results 
of~\cite{gz,bz,gz2}.  Consider a Lagrangian which is a function of $n$ real 
field strengths $F^a_{\mu\nu}$ and of some other fields $\chi^i$ and their 
derivatives $\chi^i_\mu = \pp_\mu\chi^i$:
\be L = L\(F^a,\chi^i,\chi^i_\mu\).\label{g2} \ee 
Since 
\be F^a_{\mu\nu} = \pp_\mu A^a_\nu - \pp_\nu A^a_\mu, \label{g3} \ee
we have the Bianchi identities
\be \pp^\mu\tF^a_{\mu\nu} = 0.\label{g4} \ee
On the other hand, if we define 
\be \tG^a_{\mu\nu} = {1\over2}\epsilon_{\mu\nu\lambda\sigma}G^{a\lambda\sigma}
\equiv 2 {\pp L\over\pp F_a^{\mu\nu}},\label{g5} \ee
we have the equations of motion
\be \pp^\mu\tG^a_{\mu\nu} = 0.\label{g6} \ee
We consider an infinitesimal transformation of the form
\bea \delta\pmatrix{G\cr F\cr} &=& 
\pmatrix{A & B\cr C & D\cr}\pmatrix{G\cr F\cr},
\label{g7} \\ \delta\chi^i &=& \xi^i(\chi),\label{g8} \eea
where $A,B,C,D$ are real $n\times n$ constant infinitesimal matrices and 
$\xi^i(\chi)$ functions of the fields $\chi^i$ (but not of their derivatives),
and ask under what circumstances the system of the equations of motion 
(\ref{g4}) and (\ref{g6}), as well as the equation of motion for the fields 
$\chi^i$ are invariant.  The analysis of~\cite{gz} shows that this is true if 
the matrices satisfy
\be A^T = -D,\quad B^T = B,\quad C^T = C,\label{g9} \ee
(where the superscript $T$ denotes the transposed matrix) and in addition the
Lagrangian changes under (\ref{g7}) and (\ref{g8}) as 
\be \delta L = {1\over4}\(FB\tF + GC\tG\).\label{g10} \ee
The relations (\ref{g9}) show that (\ref{g7}) is an infinitesimal
transformation of the real noncompact symplectic group $Sp(2n,R)$ which has 
$U(n)$ as maximal compact subgroup.  The finite form is
\be \pmatrix{G'\cr F'\cr} = \pmatrix{a & b\cr c & d\cr}\pmatrix{G\cr F\cr},
\label{g11} \ee
where the $n\times n$ real submatrices satisfy
\be c^Ta = a^Tc,\quad b^Td = d^Tb, \quad d^Ta - b^Tc = 1.\label{g12} \ee
For the $U(n)$ subgroup, one has in addition
\be A = D,\quad B = -C, \label{n0} \ee
or, in finite form, 
\be a = d,\quad b = -c. \label{n1} \ee

Notice that the Lagrangian is not invariant.  In~\cite{gz} we showed, however,
that the derivative of the Lagrangian with respect to an invariant parameter
{\it is} invariant.  The invariant parameter could be a coupling constant or an
external background field, such as the gravitational field, which does not 
change under duality rotations.  It follows that the energy-momentum tensor,
which can be obtained as the variational derivative of the Lagrangian with
respect to the gravitational field, is invariant under duality rotations.  No
explicit check of its invariance, as was done in~\cite{gib,gib2,bi,sch}, 
is necessary.  Using (\ref{g7}) and (\ref{g9}) it is easy to verify that 
\be \delta\(L - {1\over4}F\tG\) = \delta L - {1\over4}\(FB\tF + GC\tG\)
\label{n2}, \ee
so (\ref{g10}) is {\it equivalent} to the invariance of $L - {1\over4}F\tG$.

The symplectic transformation (\ref{g11}) can be written in a complex basis as 
\be \pmatrix{F'+iG'\cr F'-iG'\cr} = 
\pmatrix{\phi_0&\phi_1^*\cr\phi_1&\phi_0^*\cr}
\pmatrix{F+iG\cr F-iG\cr},\label{g13} \ee
where $*$ means complex conjugation and the submatrices satisfy
\be \phi_0^T\phi_1 = \phi_1^T\phi_0, \quad
\phi_0^{\dag}\phi_0 - \phi_1^{\dag}\phi_1 =1.\label{g14} \ee
The relation between the real and the complex basis is
\bea \;\;2a &= \phi_0 + \phi^*_0 - \phi_1 - \phi^*_1, \quad
2ib &= \phi_0 - \phi^*_0 - \phi_1 + \phi^*_1, \nonumber \\
-2ic &= \phi_0 - \phi^*_0 + \phi_1 - \phi^*_1, \quad
\;2d &= \phi_0 +\phi^*_0 + \phi_1 + \phi^*_1. \label{g15} \eea
In~\cite{gz,bz} we also described scalar fields valued in the quotient space
$Sp(2n,R)/U(n)$.  The quotient space can be parameterized by a complex symmetric
$n\times n$ matrix $K=K^T$ whose real part has positive eigenvalues, or
equivalently by a complex symmetric matrix $Z = Z^T$ such that $Z^{\dag}Z$ has
eigenvalues smaller than 1.  They are related by
\be K = {1-Z^*\over1+Z^*}, \quad Z = {1-K^*\over1+K^*}.\label{g16} \ee
These formulae are the generalization of the well-known map between the
Lobachevski\u{\i} unit disk and the Poincar\'e upper half-plane: $Z$ corresponds
to the single complex variable parameterizing the unit disk, $iK$ to the one
parameterizing the upper half plane.

Under $Sp(2n,R)$
\be K\to K' = \(-ib + aK\)\(d + icK\)^{-1}, \quad
Z\to Z' = \(\phi_1 + \phi_0^*Z\)\(\phi_0 + \phi_1^*Z\)^{-1}, \label{g17} \ee
or, infinitesimally, 
\be \delta K = -iB + AK - KD -iKCK, \quad \delta Z = V + T^*Z - ZT -iZV^*Z, 
\label{g18} \ee
where 
\be T = - T^{\dag}, \quad V = V^T.\label{g19} \ee

The invariant nonlinear kinetic term for the scalar fields can be obtained from
the K\"ahler metric~\cite{bg2}
\be \Tr\(dK^*{1\over K + K^*}dK{1\over K + K^*}\) = 
\Tr\(dZ{1\over1-Z^*Z}dZ^*{1\over1-ZZ^*}\) \label{g20} \ee
which follows from the K\"ahler potential
\be \Tr\ln\(1-ZZ^*\)\quad {\rm or} \quad \Tr\ln(K + K^*),\label{21} \ee
which are equivalent up to a K\"ahler transformation.
It is not hard to show that the metric (\ref{g20}) is positive definite.
In this section the normalization of the fields $F_{\mu\nu}^a$ has been chosen
to be canonical when $iK$ is set equal to the unit matrix, {\it i.e.}, when the
self-duality group reduces to the $U(n)$ subgroup; the full $Sp(2n,R)$ 
self-duality can be realized when the matrix $K$ is a function of scalar fields.
Throughout this paper we assume a flat background space-time metric; the
generalization to a nonvanishing gravitational field is 
straightforward~\cite{gz}--\cite{gib2}.
\section{Born-Infeld theory}
\setcounter{equation}{0}
As a particularly simple example we consider the case when there 
is only one 
tensor $F_{\mu\nu}$ and no additional fields.  Our equations become
\be \tG = 2{\pp L\over\pp F}, \label{b1} \ee
\be \delta F = \lambda G, \quad \delta G = - \lambda F \label{b2}\ee
and
\be \delta L = {1\over4}\lambda\(G\tG - F\tF\).\label{b3}\ee
We have restricted the duality transformation to the compact subgroup
$U(1)\cong SO(2)$, as appropriate when no additional fields are present.  So
$A=D=0,\;C= -B=\lambda.$

Since $L$ is a function of $F$ alone, we can also write 
\be \delta L = \delta F{\pp L\over\pp F} = \lambda G{1\over2}\tG. \label{b4}\ee
Comparing (\ref{b3}) and (\ref{b4}), which must agree, we find 
\be G\tG + F\tF = 0.\label{b5}\ee
Together with (\ref{b1}), this is a partial differential equation for $L(F)$, 
which is the condition for the theory to be duality invariant.  
If we introduce the complex field
\be M = F-iG,\label{b6}\ee
(\ref{b5}) can also be written as
\be M\tM^* = 0.\label{b7}\ee

    Clearly, Maxwell's theory in vacuum satisfies (\ref{b5}), or (\ref{b7}), as
expected.  A more interesting example is the Born-Infeld theory~\cite{bi}, 
given by the Lagrangian
\be L = {1\over g^2}\(-\Delta^{1\over2} + 1\), \label{b8}\ee
where
\be \Delta = -\det\(\eta_{\mu\nu} + gF_{\mu\nu}\) = 1 + {1\over2}g^2F^2 -
g^4\({1\over4}F\tF\)^2.\label{b9}\ee
For small values of the coupling constant $g$ (or for weak fields) $L$
approaches the Maxwell Lagrangian.  We shall use the abbreviation 
\be \beta = {1\over4}F\tF.\label{b10}\ee
Then
\be {\pp\Delta\over\pp F} = g^2F - \beta g^4\tF,\label{b11}\ee
\be \tG = 2{\pp L\over\pp F} = 
-\Delta^{-{1\over2}}\(F - \beta g^2\tF\),\label{b12}\ee
and
\be G = \Delta^{-{1\over2}}\(\tF + \beta g^2F\).\label{b13}\ee
Using (\ref{b12}) and (\ref{b13}), it is very easy to check that $G\tG =
-F\tF$: the Born-Infeld theory is duality invariant.  It is also not too
difficult to check that $\pp L/\pp g^2$ is actually {\it invariant} under
(\ref{b2}) and the same applies to $L - {1\over4}F\tG$ (which in this case turns
out to be equal to $-g^2\pp L/\pp g^2$).  These invariances are expected from 
our general theory.

It is natural to ask oneself whether the Born-Infeld theory is the most general
physically acceptable solution of (\ref{b5}).  This was investigated in 
\cite{gib} where a negative result was reached: more general Lagrangians 
satisfy (\ref{b5}), the arbitrariness depending on a function of one variable.
We discuss this in detail in Section 6.

\section{Schr\"odinger's formulation of Born's theory}
\setcounter{equation}{0}
Schr\"odinger~\cite{sch} noticed that, for the Born-Infeld theory (\ref{b8}), 
$F$ and $G$ satisfy not only (\ref{b5}) [or (\ref{b7})], but also the more 
restrictive relation
\be M\(M\tM\) - \tM M^2 = {g^2\over8}\tM^*\(M\tM\)^2.\label{s1}\ee
We have verified this by an explicit, although lengthy, calculation using
(\ref{b6}), (\ref{b12}), (\ref{b13}) and (\ref{b9}). Schr\"odinger did not give
the details of the calculation, presenting instead convincing arguments based on
particular choices of reference systems.  One can write (\ref{s1}) as
\be {\pp \L\over\pp M} = g^2\tM^*,\label{s2}\ee
where
\be \L = 4{M^2\over\(M\tM\)}, \label{s3}\ee
and Schr\"odinger proposed $\L$ as the Lagrangian of the theory, instead of 
(\ref{b8}).  Of course, $\L$ is a Lagrangian in a different sense than $L$,
which is a field Lagrangian in the usual sense.  Multiplying (\ref{s1}) by $M$
and saturating the unwritten indices $\mu\nu$, the left hand side vanishes, so
that (\ref{b7}) follows.  Using~(\ref{s1}) it is easy to see that $\L$ is 
pure imaginary: $\L = - \L^*.$
Schr\"odinger also pointed out that, if we introduce a map
\be {1\over g^2}{\pp\L\over\pp M} = f(M),\label{s5}\ee
so that 
(\ref{s1}) or (\ref{s2}) can be written as
\be f(M) = \tM^*,\label{s6}\ee
the square of the map is the identity map
\be f\(f(M)\) = M.\label{s7}\ee
This, together with the properties
\be f(\tM) = -{\tilde f}(M), \quad f(\M) = f(M)^*,\label{s8}\ee
ensures the consistency of (\ref{s1}).  Schr\"odinger used the Lagrangian
(\ref{s3}) to construct a conserved, symmetric energy-momentum tensor.  We have
checked that, when suitably normalized, his energy-momentum tensor agrees with
that of Born and Infeld up to an additive term proportional to $\eta_{\mu\nu}$.

Schr\"odinger's formulation is very clever and elegant and it has the advantage
of being {\it manifestly} covariant under the duality rotation 
$M\to Me^{i\lambda}$
which is the finite form of (\ref{b2}). It is also likely that, as he seems to
imply, his formulation is fully equivalent to the Born-Infeld theory
(\ref{b8}), which would mean that the more restrictive equation (\ref{s1})
eliminates the remaining ambiguity in the solutions of (\ref{b7}).  This virtue
could actually be a weakness if one is looking for more general duality 
invariant theories.

\section{General solution of the self-duality equation}
\setcounter{equation}{0}

The self-duality equation (\ref{b5}) can be solved in general as follows.
Assuming Lorentz invariance in four dimensional space-time, the Lagrangian must
be a function of the two invariants
\be \alpha = {1\over4}F^2, \quad \beta = {1\over4}F\tF, \quad L = L(\alpha,
\beta). \label{gs1}\ee
Now 
\be \tG = 2{\pp L\over\pp F} = L_\alpha F + L_\beta \tF, \quad G = - L_\alpha\tF
+ L_\beta F, \label{gs2}\ee
where we have used the standard notation $L_\alpha = \pp L/\pp\alpha,\;
L_\beta = \pp L/\pp\beta.$  Substituting these expressions in (\ref{b5}) we
obtain
\be \[\(L_\beta\)^2 - \(L_\alpha\)^2 + 1\]\beta + 2L_\alpha L_\beta\alpha = 0.
\label{gs3}\ee
This partial differential equation for $L$ can be simplified by the change of
variables
\be x = \alpha, \quad y = \(\alpha^2 + \beta^2\)^{1\over2}, \label{gs4}\ee
which gives 
\be \(L_x\)^2 - \(L_y\)^2 = 1. \label{gs5}\ee
Alternatively one can use the variables
\be p = {1\over2}(x+y),\quad q = {1\over2}(x-y),\label{gs6}\ee
to obtain the form
\be L_pL_q = 1. \label{gs7}\ee

The equation (\ref{gs5}), or (\ref{gs7}), has been studied extensively in
mathematics and there are several methods to obtain its general 
solution~\cite{cour}.  (It is interesting that the same equation occurs in a
study of 5-dimensional Born-Infeld theory~\cite{perry}.)  In our case we must
also impose the physical boundary condition that the Lagrangian should
approximate the Maxwell Lagrangian
\be L_M = -\alpha = - x = - p - q \label{gs8}\ee
when the field strength $F$ is small.

According to one of the methods given in Courant-Hilbert, the general solution
of (\ref{gs7}) is given by 
\bea L &=& {2p\over v'(s)} + v(s), \label{gs9}\\
q &=& {p\over[v'(s)]^2} + s, \label{gs10} \eea
where the arbitrary function $v(s)$ is determined by the initial values:
\bea L(p = 0,q) &=& v(q), \label{gs11}\\
L_p(p = 0,q) &=& {1\over v'(q)}. \label{gs12} \eea
One must solve for $s(p,q)$ from (\ref{gs10}) and substitute into (\ref{gs9}).
To verify~\cite{perry} that these equations solve (\ref{gs7}), differentiate 
(\ref{gs9}) and (\ref{gs10}):
\bea dL &=& {2dp\over v'} + \(v'  - {2p\over[v']^2}v''\)ds, \label{gs13}\\
dq &=& {dp\over v'^2} + \(1 - {2p\over[v']^3}v''\)ds, \label{gs14} \eea
and eliminate $ds$ between (\ref{gs13}) and (\ref{gs14}) to obtain
\be dL = {1\over v'}dp + v'dq, \label{gs15}\ee
{\it i.e.},
\be L_p = {1\over v'}, \quad L_q = v', \quad L_pL_q = 1. \label{gs16}\ee

The condition that $L$ should approach the Maxwell
Lagrangian for small field strengths implies that
\be v(s) = L(p=0,s)\cong -s\label{gs24}\ee
for small $s$.  

It is trivial to check the above procedure for the Maxwell Lagrangian, and we
shall not do it here.  The Born-Infeld Lagrangian (with $g=1$ for simplicity) is
given by 
\bea L_{BI} &=& - \Delta^{1\over2} + 1, \label{gs17}\\
\Delta &=& (1 + 2p)(1+ 2q), \label{gs18} \eea
in terms of the variables $p$ and $q$.  Setting $p =0$ we see that this
corresponds to 
\bea v(s) &=& -(1+2s)^{1\over2} + 1,\label{gs19}\\
v'(s) &=& -(1+2s)^{-{1\over2}} .\label{gs20}\eea
Then (\ref{gs10}) gives
\be q = p(1+2s) + s, \label{gs21}\ee
which is solved by 
\be s = {q-p\over 1 + 2p}, \quad 1 + 2s = {1 + 2q\over 1 + 2p}.\label{gs22}\ee
Using (\ref{gs9}), we reconstruct the Lagrangian
\be L_{BI} = -2p\({1 + 2q\over1 + 2p}\)^{1\over2} - 
\({1 + 2q\over1 + 2p}\)^{1\over2} + 1 = - \[(1 + 2p)(1 + 2q)\]^{1\over2} + 1.
\label{gs23}\ee
Unfortunately, in spite of this elegant method for finding
solutions of the self-duality equation, it seems very difficult to find new
explicit solutions given in terms of simple functions.  The reason is that,
even for a simple function $v(s)$, solving the equation (\ref{gs10}) for $s$
gives complicated functions $s(p,q)$.

\section{Axion, dilaton and $SL(2,R)$}
\setcounter{equation}{0}
It is well known that, if there are additional scalar fields which transform
nonlinearly, the compact group duality invariance can be enhanced to a duality
invariance under a larger noncompact group (see, {\it e.g.},~\cite{gz} and 
references
therein).  In the case of the Born-Infeld theory, just as for Maxwell's theory,
one complex scalar field suffices to enhance the $U(1)\cong SO(2)$ invariance to
the $SU(1,1)\cong SL(2,R)$ noncompact duality invariance.  This is pointed out
in~\cite{gib2}, but it also follows from the considerations of our 
paper~\cite{gz}.  In the example under consideration, $K$ is a single complex 
field, not an $n\times n$ matrix.  In order to agree with today's more 
standard notation we shall use 
\be S = iK = S_1 + iS_2 = a + ie^{-\phi}, \quad S_2 > 0,\label{d1}\ee
where $\phi$ is the dilaton and $a$ is the axion.  For $SL(2,R)\cong Sp(2,R)$, 
the matrices $A,B,C,D$ are real numbers and $A=-D,\;B$ and $C$ are independent.
Then the infinitesimal $SL(2,R)$ transformation is
\be \delta S = B + 2A S - C S^2, \label{d2}\ee
and the finite transformation is 
\be S' = {aS + b\over cS +d}, \quad ad-bc = 1. \label{n3}\ee
For the $SO(2)\cong U(1)$ subgroup, $A=0,\;C= -B = \lambda,$
\be \delta S = -\lambda - \lambda S^2 .\label{d3}\ee
The scalar kinetic term, proportional to 
\be {\pp_\mu\S\pp^\mu S\over(S-\S)^2},\label{d4}\ee
is invariant under the nonlinear transformation (\ref{d2}) which, in terms of
$S_1,S_2$, takes the form 
\be \delta S_1 = B+ 2A S_1 - C\(S_1^2 - S_2^2\) , \quad
\delta S_2 = 2A S_2 - 2C S_1S_2.\label{d5}\ee
Since the scalar kinetic term is separately invariant, we assume from now on
that $\hL(S,F)$ does not depend on the derivatives of $S$.

The full noncompact duality transformation on $F_{\mu\nu}$ is now
\be \delta G = AG + BF, \quad \delta F = CG + DF, \quad D = -A,\label{d6}\ee
and we are seeking a Lagrangian $\hL(S,F)$ which satisfies
\be \delta \hL = {1\over4}\(FB\tF + GC\tG\),\label{d7}\ee
where
\be \delta\hL = \delta F{\pp\hL\over\pp F} + \delta S_1{\pp\hL\over\pp S_1} +
\delta S_2{\pp\hL\over\pp S_2} ,\label{d8}\ee
and now
\be \tG = 2{\pp\hL\over\pp F} .\label{dd}\ee
Equating (\ref{d7}) and (\ref{d8}) we see that $\hL$ must satisfy
\be {1\over4}\(CG\tG - BF\tF\) - {1\over2}AF\tG + 
\delta S_1{\pp\hL\over\pp S_1} + \delta S_2{\pp\hL\over\pp S_2} =0.
\label{d9}\ee

This equation can be solved as follows.  Assume that $L(\F)$ satisfies
(\ref{b1}) and (\ref{b5}), {\it i.e.}
\be \G\tcG + \F\tcF = 0,\label{d10}\ee
where
\be \tcG = 2{\pp L\over\pp \F}. \label{d11} \ee
For instance, the Born-Infeld Lagrangian $L(\F)$ does this.  Then 
\be\hL(S,F) = L(S_2^{1\over2}F) + {1\over4}S_1F\tF\label{d12}\ee
satisfies (\ref{d9}).  Indeed
\be {\pp\hL(S,F)\over\pp F} = {\pp L\over\pp\F}S_2^{1\over2} + {1\over2}S_1\tF.
\label{d13}\ee
So 
\be \tG = \tcG S_2^{1\over2} + S_1\tF, \label{d14}\ee
\be G = \G S_2^{1\over2} + S_1 F,\label{d15}\ee
where we have defined 
\be \F = S_2^{1\over2} F, \label{d16}\ee
and $\tcG$ is given by (\ref{d11}).  Now
\be G\tG = \G\tcG S_2 + S_1^2F\tF + 2S_1\F\tcG. \label{d17}\ee
Using (\ref{d10}) in this equation we find
\be G\tG = \(S^2_1 - S_2^2\)F\tF + 2S_1\F\tcG. \label{d18}\ee
We also have
\be F\tG = \F\tcG + S_1F\tF. \label{d19}\ee
On the other hand, since
\be {\pp L\over\pp S_2^{1\over2}} = {\pp L\over\pp\F}F = {1\over2}\tcG F,
\label{d20}\ee
we obtain
\be {\pp\hL\over\pp S_2} = {\pp L\over\pp S_2^{1\over2}}{1\over2}
S_2^{-{1\over2}} =
{1\over4}\tcG S_2^{-{1\over2}}F = {1\over4}\tcG\F S_2^{-1}. \label{d21}\ee
In addition 
\be  {\pp\hL\over\pp S_1} = {1\over4}F\tF.\label{d22}\ee
Using (\ref{d18}), (\ref{d19}), (\ref{d21}) and (\ref{d22}), together with 
(\ref{d5}), we see that (\ref{d9}) is satisfied.
It is easy to check that the scale invariant combinations $\F$ and $\G$, given 
by (\ref{d16}) and (\ref{d11}) have the very simple transformation law
\be \delta\F = S_2C\G, \quad \delta\G = - S_2C\F, \label{d23}\ee
{\it i.e.,}  they transform according to the $U(1)\cong SO(2)$ compact subgroup
just as $F$ and $G$ in (\ref{b2}), but with the parameter $\lambda$ replaced by
$S_2C$.  If $L(\F)$ is the Born-Infeld Lagrangian, the theory with scalar fields
given by $\hL$ in (\ref{d12}) can also be reformulated \`a la Schr\"odinger.
>From (\ref{d15}) and (\ref{d16}) solve for $\F$ and $\G$ in terms of $F,G,S_1$  
and $S_2$.   Then $\cM = \F -i\G$ must satisfy the same equation (\ref{s1}) that
$M$ does when no scalar fields are present.

\section{Duality as a Legendre transformation}
\setcounter{equation}{0}

We have observed that, even in the general case of $Sp(2n,R)$, although the
Lagrangian is not invariant, the combination [see (\ref{n2})]
\be \hL - {1\over4}F\tG \label{dl1}\ee 
is invariant.  Here we restrict ourselves to the case of $SL(2,R)$, one tensor
$F_{\mu\nu}$ and one complex scalar field $S = S_1 + iS_2$.  As in Section 5, we
use the notation $\hL$ to denote the part of the Lagrangian that depends on 
the scalar fields, as well as on $F_{\mu\nu}$, but not on scalar derivatives. 
Then
\be \hL(S_1,S_2,F) -  {1\over4}F\tG = \hL(S'_1,S'_2,F') - {1\over4}F'\tG',
\label{dl2}\ee
where
\bea \pmatrix{G'\cr F'\cr} &=& \pmatrix{a & b\cr c & d\cr}\pmatrix{G\cr F\cr}, 
\quad S' = {aS +b\over cS+d}, \quad ad -bd = 1,\label{dl3}\\
\tG &=& 2{\pp\hL\over\pp F}. \label{dl4}\eea

There are several interesting special cases of this invariance statement.  The
first corresponds to $a=d=1,\; c=0,\; b$ arbitrary, which gives 
\be G' = G + bF, \quad F' = F, \quad S'_1 = S_1 + b, \quad S'_2 = S_2.
\label{dl5}\ee
The second corresponds to $b=c=0,\;d= 1/a,\; a$ arbitrary, which gives
\be G' = aG, \quad F' = {1\over a}F, \quad S' = a^2S,
\quad S'_1 = a^2S_1, \quad S'_2 = a^2S_2 . \label{d1s}\ee
The third corresponds to $a=d=0,\;b= -1/c,\; c$ arbitrary, which gives
\be G' = - {1\over c}F, \quad F' = cG, \quad S' = -{1\over c^2S},
\quad S'_1 = - {S_1\over c^2|S|^2}, \quad S'_2 = {S_2\over c^2|S|^2}.
\label{dl7}\ee
Using (\ref{dl5}) in (\ref{dl2}) we find 
\be \hL(S_1,S_2,F) -  {1\over4}F\tG = \hL(S_1 + b,S_2,F) - {1\over4}F
\(\tG + b\tF\).  \label{dl8}\ee
Taking $b = - S_1$, we obtain
\be \hL(S_1,S_2,F) = \hL(0,S_2,F) + {1\over4}S_1F\tF,  \label{dl9}\ee
which gives the dependence of $\hL$ on $S_1$, in agreement with (\ref{d12}).
This choice for the constant $b$ is allowed because this part of the
Lagrangian, which does not include the kinetic term for the scalar fields, does
not contain derivatives of the scalar fields.
Using (\ref{d1s}) in (\ref{dl2}) we find 
\be \hL(S_1,S_2,F) - {1\over4}F\tG = \hL\(a^2S_1,a^2S_2,{1\over a}F\) - 
{1\over4}F\tG, \label{d3s}\ee {\it i.e.},
\be \hL(S_1,S_2,F) = \hL\(a^2S_1,a^2S_2,{1\over a}F\). \label{d4s}\ee
Setting $S_2=0$ in this equation, we see that $\hL(S_1,0,F)$ is a function of 
$S_1^{1\over2}F$, in agreement with the more precise statement (\ref{dl9}). 
Setting instead $S_1=0$, we   
find that $\hL(0,S_2,F)$ is a function of $S_2^{1\over2}F$, in agreement with 
(\ref{d12}). 

Using (\ref{dl7}) in (\ref{dl2}) we find
\be \hL(S_1,S_2,F) -  {1\over4}F\tG = \hL\(-{S_1\over c^2|S|^2},
{S_2\over c^2|S|^2},cG\) + {1\over4}G\tF,  \label{dl10}\ee
{\it i.e.},
\be \hL\(-{S_1\over c^2|S|^2},{S_2\over c^2|S|^2},cG\) = 
\hL(S_1,S_2,F) -  {1\over2}F\tG, \label{dl11}\ee
or
\be \hL\(-{1\over c^2S},cG\) = \hL(S,F) -  {1\over2}F\tG. \label{dl12}\ee

We have shown that the Ansatz (\ref{d12}) of Section 5 is a natural consequence 
of the invariance of $\hL - {1\over4}F\tG$.
Equation (\ref{dl12}) with (\ref{dl4}) can be interpreted as a Legendre
transformation.  Given a Lagrangian $\hL(S,F)$, {\it define} the dual 
Lagrangian $\hL_D(S,F_D)$, a function of the dual field $F_D$, by 
\be \hL_D(S,F_D) + \hL(S,F) = {1\over2}FF_D, \label{dl21}\ee
\be F_D = 2{\pp\hL\over\pp F}, \quad F = 
2{\pp \hL_D\over\pp F_D}.\label{dl22}\ee
With these definitions, the dual of the dual of a function equals the original
function.\footnote{The unconventional factor 1/2 on the right hand side of
(\ref{dl21}) is introduced to avoid overcounting when summing over the indices
of the antisymmetric tensors $F$ and $F_D$.}
In general, the dual Lagrangian is a very different function from the original
Lagrangian.  For a self-dual theory, if we set 
\be F_D = \tG,\quad \tF_D = - G,\label{dl23}\ee
we see from (\ref{dl12}) that
\be - \hL\(-{1\over c^2S},cG\) = \hL_D(S,\tG), \label{dl24}\ee
which must be independent of $c$, since $G$ is.  

The above argument can be inverted.  Let the Legendre transformation 
(\ref{dl21}) produce a dual Lagrangian given by (\ref{dl24}) with $c=1$, or
\be \hL_D(S,F_D) = - \hL\(-{1\over S},-\tF_D\) = - \hL\(-{1\over S},G\)
. \label{ll1} \ee
It then follows that $\hL - {1\over4}F\tG$ is invariant under (\ref{dl7}) with 
$c=1$, {\it i.e.}, 
\be G' = - F, \quad F' = G, \quad S' = - {1\over S}. \label{ll2}\ee
If we now assume that it is also invariant under (\ref{dl5}) with arbitrary
$b$, it follows that it is invariant under the entire group $SL(2,R)$.  Indeed,
if we call $t_b$ the transformation (\ref{dl5}) and $s$ the transformation
(\ref{ll2}), the product $t_bst_{b'}st_{b''}$ gives the most general
transformation of $SL(2,R)$.

If we normalize the scalar field differently, taking {\it e.g.}, instead of $S$,
\be \tau = cS,\quad L'(\tau,F) = \hL(S,F), \label{dt20}\ee
\be L'_D(\tau,F'_D) + L'(\tau,F) = {1\over2c}FF'_D, \label{dt21}\ee
and write the Legendre transformation as 
\be 2{\pp L'(\tau,F)\over\pp F} = {1\over c}F'_D, 
\quad 2{\pp L'_D(\tau,F'_D)\over\pp F'_D} = {1\over c}F,\label{dt22}\ee
we see that
\be F'_D = cF_D = c \tG, \label{dt23}\ee
and 
\bea L'_D(\tau,F'_D) &=& \hL_D(S,F_D) = - \hL\(-{1\over c^2S},-c\tF_D\)
\nonumber \\ &=& - L'\(-{1\over cS},-c\tF_D\) = - L'\(-{1\over\tau},-\tF'_D\), 
\label{dt24}\eea
for a self-dual theory.

A standard normalization~\cite{wit,seib} is $c=4\pi$, in which case the
expectation value of the field $\tau$ is
\be <\tau> = {\theta\over2\pi} + i{4\pi\over g^2}. \label{dt25}\ee
In the presence of magnetically charged particles and dyons (both electrically
and magnetically charged) the invariance of the charge lattice
restricts~\cite{olive} the $SL(2,R)$ group to 
the $SL(2,Z)$ subgroup generated by
\be \tau\to -{1\over\tau},\quad \tau \to \tau + 1.\label{last}\ee
At the quantum
level the Legendre transformation corresponds to the integration over the field
$F$ in the functional integral, after adding to the Lagrangian $\hL$ a term
$-{1\over2}FF_D$.

\section{Concluding remarks}
\setcounter{equation}{0}

Nonlinear electromagnetic Lagrangians, like the Born-Infeld Lagrangian, can be
supersymmetrized~\cite{des,cec} by means of the four-dimensional $N=1$
superfield formalism, and this can be done even in the presence of supergravity.
When the Lagrangian is self-dual, it is natural to ask whether its
supersymmetric extension possesses a self-duality property that can be 
formulated
in a supersymmetric way.  We were not able to do this in the nonlinear case.
When the Lagrangian is quadratic in the fields $F^a_{\mu\nu}$, the problem as
been solved in~\cite{bg}, where the combined requirements of supersymmetry and
self-duality were used to constrain the form or the weak coupling 
($S_2\to\infty$) limit of the effective Lagrangian from string theory,
in which one neglects the nonabelian nature of the gauge fields.

The $SL(2,Z)$ subgroup of $SL(2,R)$ that is generated by the 
elements  $4\pi S\to -1/4\pi S$ and $S\to S + 1/4\pi$ relates different 
string theories~\cite{ssdw} to one another. 

The generalization of~\cite{gz} to two dimensional theories~\cite{fer} has 
been used to derive the K\"ahler potential for moduli and matter fields in
effective field theories from superstrings.  In this case the scalars are valued
on a coset space ${\cal K}/{\cal H},\; {\cal K}\in SO(n,n),\; {\cal H}\in
SO(n)\times SO(n).$ The kinetic energy is invariant under ${\cal K}$, and the
full classical theory is invariant under a subgroup of ${\cal K}$.
String loop corrections reduces the invariance to a discrete
subgroup that contains the $SL(2,Z)$ group generated by $T\to 1/T,\; T\to T -
i$, where Re$T$ is the squared radius of compactification in string units.

\vskip .2in
\noindent{\bf Acknowledgements.} We are grateful for the hospitality provided by
the Isaac Newton Institute where this work was initiated.  We thank Gary 
Gibbons, David Jackson,
Bogdan Morariu, David Olive, Harold Steinacker, Kelly Stelle and 
Peter West for inspiring conversations.  This work was supported in part by the
Director, Office of Energy Research, Office of High Energy and Nuclear Physics,
Division of High Energy Physics of the U.S. Department of Energy under Contract
DE-AC03-76SF00098 and in part by the National Science Foundation under grant
PHY-95-14797.  

\end{document}